\documentclass[twocolumn,english,aps,prl,showpacs,superscriptaddress,nobibnotes,longbibliography]{revtex4-1}
\usepackage[T1]{fontenc}
\usepackage[latin9]{inputenc}
\usepackage{units}
\usepackage{amsmath}
\usepackage{amssymb}
\usepackage{graphicx}

\makeatletter
\usepackage[breaklinks]{hyperref}
\usepackage{breakurl}

\makeatother

\usepackage{babel}
\begin{document}
\title{Composite-fringe atom interferometry for high dynamic-range sensing}
\author{Chen Avinadav}
\thanks{These authors contributed equally to this work.~\\
chen.avinadav@weizmann.ac.il~\\
dimitry.yankelev@weizmann.ac.il}
\affiliation{Department of Physics of Complex Systems, Weizmann Institute of Science,
Rehovot 7610001, Israel}
\affiliation{Rafael Ltd, Haifa 3102102, Israel}
\author{Dimitry Yankelev}
\thanks{These authors contributed equally to this work.~\\
chen.avinadav@weizmann.ac.il~\\
dimitry.yankelev@weizmann.ac.il}
\affiliation{Department of Physics of Complex Systems, Weizmann Institute of Science,
Rehovot 7610001, Israel}
\affiliation{Rafael Ltd, Haifa 3102102, Israel}
\author{Ofer Firstenberg}
\affiliation{Department of Physics of Complex Systems, Weizmann Institute of Science,
Rehovot 7610001, Israel}
\author{Nir Davidson}
\affiliation{Department of Physics of Complex Systems, Weizmann Institute of Science,
Rehovot 7610001, Israel}
\begin{abstract}
Atom interferometers offer excellent sensitivity to gravitational
and inertial signals but have limited dynamic range. We introduce
a scheme that improves this trade-off by a factor of 50 using composite,
moiré-like, fringes, obtained from sets of measurements with slightly
varying interrogation times. We analyze analytically the performance
gain in this approach and the trade-offs it entails between sensitivity,
dynamic range, and bandwidth, and we experimentally validate the analysis
over a wide range of parameters. Combining composite-fringe measurements
with a particle-filter estimation protocol, we demonstrate continuous
tracking of a rapidly varying signal over a span two orders of magnitude
larger than the dynamic range of a traditional atom interferometer.
\end{abstract}
\maketitle

\section{Introduction}

Atom interferometry (AI) \citep{Varenna2014} enables highly sensitive
and accurate sensing of gravitational \citep{Peters2001,Snadden1998,Sorrentino2012}
and inertial forces \citep{Barrett2014,Canuel2006,Stockton2011,Gustavson1997,Dickerson2013,Savoie2018}.
In addition to laboratory-based experiments in fundamental physics,
such as tests of general relativity \citep{Dimopoulos2007,Mueller2010,Hohensee2011,Aguilera2014,Zhou2015a}
and precision measurement of physical constants \citep{Weiss1993,Fixler2007,Bouchendira2011,Rosi2014,Parker2018},
AIs for field-applications are being developed worldwide \citep{Bongs2019}.
Mobile atomic gravimeters and gravity-gradiometers \citep{Wu2009,Bidel2013,Farah2014,Freier2016,Barrett2016,Menoret2018,Zhu2018,Becker2018,Wu2019,Lamb2019}
have been demonstrated for geophysical surveys on land \citep{Wu2009,Wu2019},
at sea \citep{Bidel2018}, and in the air \citep{Bidel2019}, while
cold atom accelerometers and gyroscopes are developed for inertial
navigation \citep{Rakholia2014,Cheiney2018,Chen2019}. Such applications
motivate the development of advanced AI techniques for robust operation
under conditions of large uncertainty and large temporal variations
in the measured signal.

As a phase-measuring instrument, a trade-off exists between an AI
sensitivity and its ambiguity-free dynamic range. The ratio of dynamic
range to sensitivity is in general fixed by the signal-to-noise ratio
(SNR), whereas the scale factor, which determines their absolute values,
may be controlled by changing the interferometer interrogation time
$T$. When prior knowledge of the measured signal is insufficient,
a standard approach involves initial measurements with low sensitivity
and high dynamic range (short $T$) and gradual progress to measurements
with high sensitivity and low dynamic range (long $T$) \citep{Wu2019}.
However, the time-averaged sensitivity per $\sqrt{\text{Hz}}$ of
this sequence is greatly reduced. Alternatively, simultaneous measurements
using two interrogation times was demonstrated in a dual-species interferometer
\citep{Bonnin2018} with an improved dynamic range of $\times5$ at
the cost of added experimental complexity. Ambiguities of AIs may
also be resolved by hybridization with classical sensors with large
dynamic range \citep{Merlet2009,Lautier2014}, especially relevant
when the measured signal varies continuously in time and substantially
changes from shot to shot. However, imperfections such as non-linearity
of the classical sensors, transfer function errors and misalignment
may limit the usefulness of this technique in harsh conditions \citep{Bidel2018},
necessitating prolonged operation at short $T$ at the expense of
sensitivity.

In this work, we introduce an approach to AI which increases its dynamic
range with minor penalty on sensitivity. We perform a set of measurements
with slightly\emph{ }varying values of $T$, corresponding to slightly
different interferometer scale factors. Together, as in a moiré effect,
these measurements constitute a composite fringe whose frequency,
as well as its phase, encodes the measured inertial signal, providing
a non-ambiguous dynamic range larger than measurements with a fixed
$T$. The increase in dynamic range scales inversely with the span
of scale factors and can reach orders of magnitude, limited only by
the experimental SNR. In addition to a static demonstration, we apply
the scheme together with a particle-filter estimator to successfully
track rapidly-varying signals, which change by more than $2\pi$ between
consecutive measurements and span hundreds of radians altogether,
while maintaining high sensitivity.

\section{Experimental setup}

\begin{figure}[t]
\begin{centering}
\includegraphics[bb=0bp 340bp 620bp 540bp,clip,width=1\columnwidth]{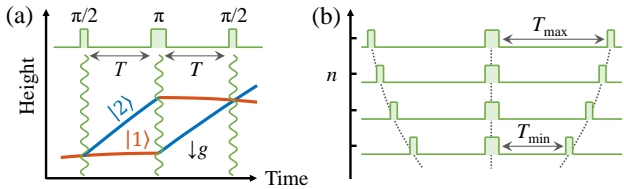}
\par\end{centering}
\centering{}\caption{(a) Schematic diagram of an atomic gravimeter, operating in a $\pi/2$-$\pi$-$\pi/2$
Mach-Zehnder geometry. Counter-propagating laser beams interact with
the atoms and generate the interferometer sequence. (b) Composite-fringe
AI comprises a set a of measurements with interrogation times $T$
that vary quadratically within a small range.\label{fig:0-apparatus-scheme}}
\end{figure}

We apply the composite fringe approach in a Mach-Zehnder AI which
measures gravitational acceleration \citep{Kasevich_1991}. A freely-falling
cold-atom ensemble interacts with pulses of counter-propagating laser
beams which stimulate two-photon transitions. A sequence of three
pulses spatially splits, redirects, and recombines the atomic wavepackets
{[}Fig.~\ref{fig:0-apparatus-scheme}(a){]}. The interferometer phase
is $\phi=\left(k_{\textrm{eff}}g-\alpha\right)T^{2}+\phi_{L}$, where
$k_{\textrm{eff}}$ is the effective atom optics two-photon wavevector,
$g$ is the gravitational acceleration, and $T$ is the time between
pulses. The relative frequency of the counter-propagating beams is
chirped at a rate $\alpha=k_{\textrm{eff}}g_{0}$, where $g_{0}$
is an approximate value of $g$, to compensate the changing Doppler
shift of the falling atoms. $\phi_{L}$ is a tunable laser phase applied
during the final $\pi/2$-pulse.

Our apparatus is described in detail in Ref. {[}\citealp{Yankelev2019}{]}.
Briefly, we trap and cool $^{87}\textrm{Rb}$ atoms and launch them
upwards using moving optical molasses. Vertical, retroreflected Raman
beams, derived from a single laser modulated at $6.834\,\textrm{GHz}$,
drive Doppler-sensitive two-photon transitions \citep{Kasevich1991a}
between $\left|F=1,m_{F}=0\right\rangle $ and $\left|F=2,m_{F}=0\right\rangle $
for state initialization and interferometry sequence. The population
fraction in $F=2$, determined by state-dependent fluorescence, constitutes
the interferometer output signal.

\section{Composite fringe analysis and performance}

The standard fringe of an AI is $S\left(\phi\right)=A-\left(C/2\right)\cos\phi$,
where $A$ and $C$ are the fringe offset and contrast. By measuring
this fringe at $N$ points (scanning either $\alpha$ or $\phi_{L}$)
at a fixed $T$, the gravitational phase $k_{\textrm{eff}}gT^{2}$
is determined with uncertainty $\sigma_{\phi}/\sqrt{N}$, with $\sigma_{\phi}$
the total phase uncertainty \emph{per shot} (see Appendix A). Therefore,
gravity is determined with uncertainty per shot $\sigma_{g,\textrm{standard}}=\sigma_{\phi}/\left(k_{\textrm{eff}}T^{2}\right)$,
over an ambiguity-free dynamic range $\Delta g_{\textrm{standard}}=2\pi/\left(k_{\textrm{eff}}T^{2}\right)$.
The ratio of dynamic range to sensitivity-per-shot is $\left(\Delta g/\sigma_{g}\right)_{\textrm{standard}}=2\pi/\sigma_{\phi}$,
depending only on the measurement phase uncertainty.

\begin{figure*}[t]
\begin{centering}
\includegraphics[clip,width=2\columnwidth]{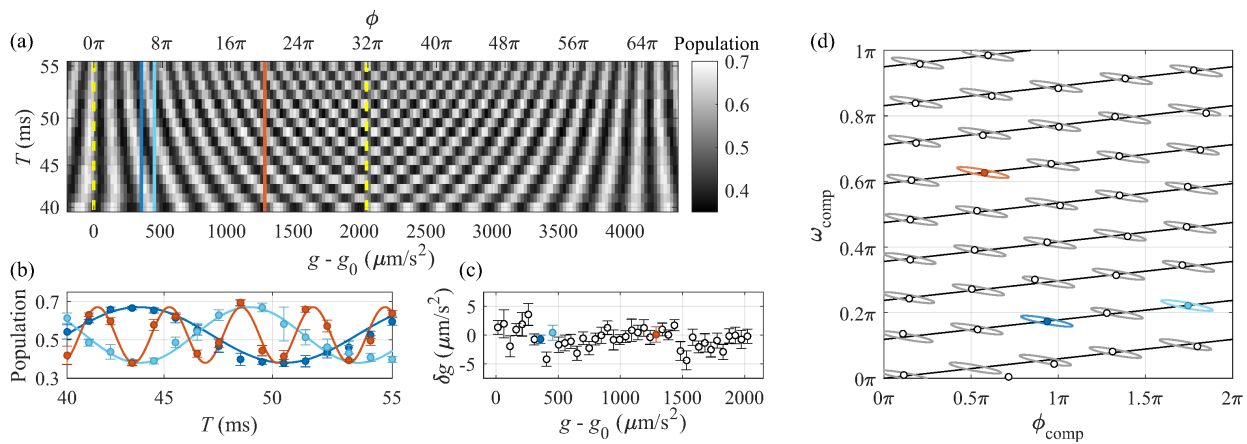}
\par\end{centering}
\centering{}\caption{Composite-fringe AI. (a) Experimental data using $N=16$ interrogation
times between $T_{\min}=\unit[40]{ms}$ and $T_{\max}=\unit[55]{ms}$.
Different gravity values are simulated by changing the Raman beams
chirp rate $\alpha=k_{\textrm{eff}}g_{0}$. The ambiguity-free dynamic
range (dashed lines) is extended by the composite-fringe approach
to $\unit[32\pi]{rad}$ at $T_{\max}$, representing a 16-fold increase
compared to standard measurements. Data is averaged over 5 shots.
(b) Composite fringes measured at different gravity values as indicated
by colored lines in (a). Proximate gravity values (dark, bright blue)
result in fringes with similar frequency but different phase, while
a larger difference in gravity (red) renders a substantial change
in frequency. Dots are measurements averaged over 5 shots, error bars
are $\pm1\sigma$, solid lines are sinusoidal fits. (c) Residual of
the gravity value fitted from each composite fringe, over the span
of one extended dynamic range. Resulting sensitivity is $\unit[1.7]{\mu m/s^{2}}$
per fringe, or $\unit[6.7]{\mu m/s^{2}}$ per shot, comparable to
$\unit[7.3]{\mu m/s^{2}}$ per shot using standard fringes at fixed
$T=T_{\max}$ in our apparatus. Residuals have mean $\unit[-0.3\pm0.4\left(2\sigma\right)]{\mu m/s^{2}}$,
consistent with no added bias. (d) Phase-frequency map of composite
fringes: circles represent the phase and frequency extracted from
fits to the fringe data in (a), solid lines are the predicted loci
of measurements within the extended dynamic range. Gravity increases
from bottom left to top right. Ellipses represent the covariance matrix
of phase and frequency estimation with the experimentally-characterized
$\sigma_{\phi}=\unit[400]{mrad}$ at 95\% confidence interval.\label{fig:1-scheme}}
\end{figure*}

Here instead, we form a composite fringe from a set of $N$ measurements
$S_{n}$, varying the scale factor $k_{\textrm{eff}}T^{2}$ linearly
by choosing variable interrogation times {[}Fig.~\ref{fig:0-apparatus-scheme}(b){]},
\begin{equation}
T_{n}^{2}=T_{\textrm{min}}^{2}+n\frac{T_{\textrm{max}}^{2}-T_{\textrm{min}}^{2}}{N-1}\,\,\,\,\,\textrm{for}\,\,\,\,n=0,\ldots,N-1,
\end{equation}
with $T_{\textrm{min}}<T_{\textrm{max}}$, keeping $\alpha$ and $\phi_{L}$
fixed. As Figs.~\ref{fig:1-scheme}(a),(b) show, the series $\left\{ S_{n}\right\} $
forms a composite fringe,\emph{ }$S_{n}=A-\left(C/2\right)\cos\left(n\omega_{\textrm{comp}}+\phi_{\textrm{comp}}\right)$,
with
\begin{align}
\phi_{\textrm{comp}} & =\left(k_{\textrm{eff}}g-\alpha\right)T_{\textrm{min}}^{2}+\phi_{L},\label{eq:phi_g}\\
\omega_{\textrm{comp}} & =\left(k_{\textrm{eff}}g-\alpha\right)\frac{T_{\textrm{max}}^{2}-T_{\textrm{min}}^{2}}{N-1}.\label{eq:w_g}
\end{align}

Unlike standard fringes with fixed $T$, here both phase and frequency
depend on $g$, albeit with very different scaling: $\phi_{\textrm{comp}}$
varies rapidly with $g$ and represents a high-resolution measurement,
similar to standard measurements with fixed $T$, whereas $\omega_{\textrm{comp}}$
varies slowly with $g$ and represents a coarse measurement {[}Fig.~\ref{fig:1-scheme}(d){]}.
This beat frequency, which is revealed by varying $T$, provides the
extended dynamic range and stands in close analogy to moiré interferometry
\citep{Oberthaler1996,Cronin2009}, while the phase of the composite
fringe is still accessible and provides the high sensitivity.

The composite-fringe frequency can be estimated unambiguously up to
$\omega_{\textrm{comp}}=\pi$, resulting in the extended dynamic range
$\Delta g_{\textrm{comp}}=\pi\left(N-1\right)/\left[k_{\textrm{eff}}\left(T_{\textrm{max}}^{2}-T_{\textrm{min}}^{2}\right)\right]$.
With respect to a fixed-$T$ measurement, the dynamic range increases
by a factor

\begin{align}
\frac{\Delta g_{\textrm{comp}}}{\Delta g_{\textrm{standard}}} & =\frac{1}{2}\frac{N-1}{1-T_{\textrm{min}}^{2}/T_{\textrm{max}}^{2}}.\label{eq:dyn-range-gain}
\end{align}
At the same time, the gravity sensitivity per shot of a composite
fringe is approximately (see Appendix A) 
\begin{equation}
\sigma_{g,\textrm{comp}}\approx\frac{\sigma_{\phi}}{k_{\textrm{eff}}T_{\min}T_{\max}}=\frac{T_{\max}}{T_{\min}}\cdot\sigma_{g,\textrm{standard}}.
\end{equation}
For $T_{\min}$ approaching $T_{\max}$, the dynamic range enhancement
is significant while the penalty in $\sigma_{g,\textrm{comp}}$ is
small. Indeed, Fig.~\ref{fig:1-scheme}(c) presents the gravity residuals
$\delta g$ obtained from fits of the measured composite fringes,
exhibiting similar sensitivity to measurements at fixed $T=T_{\max}$
with no observed systematic bias.

\begin{figure}
\begin{centering}
\includegraphics[clip,width=1\columnwidth]{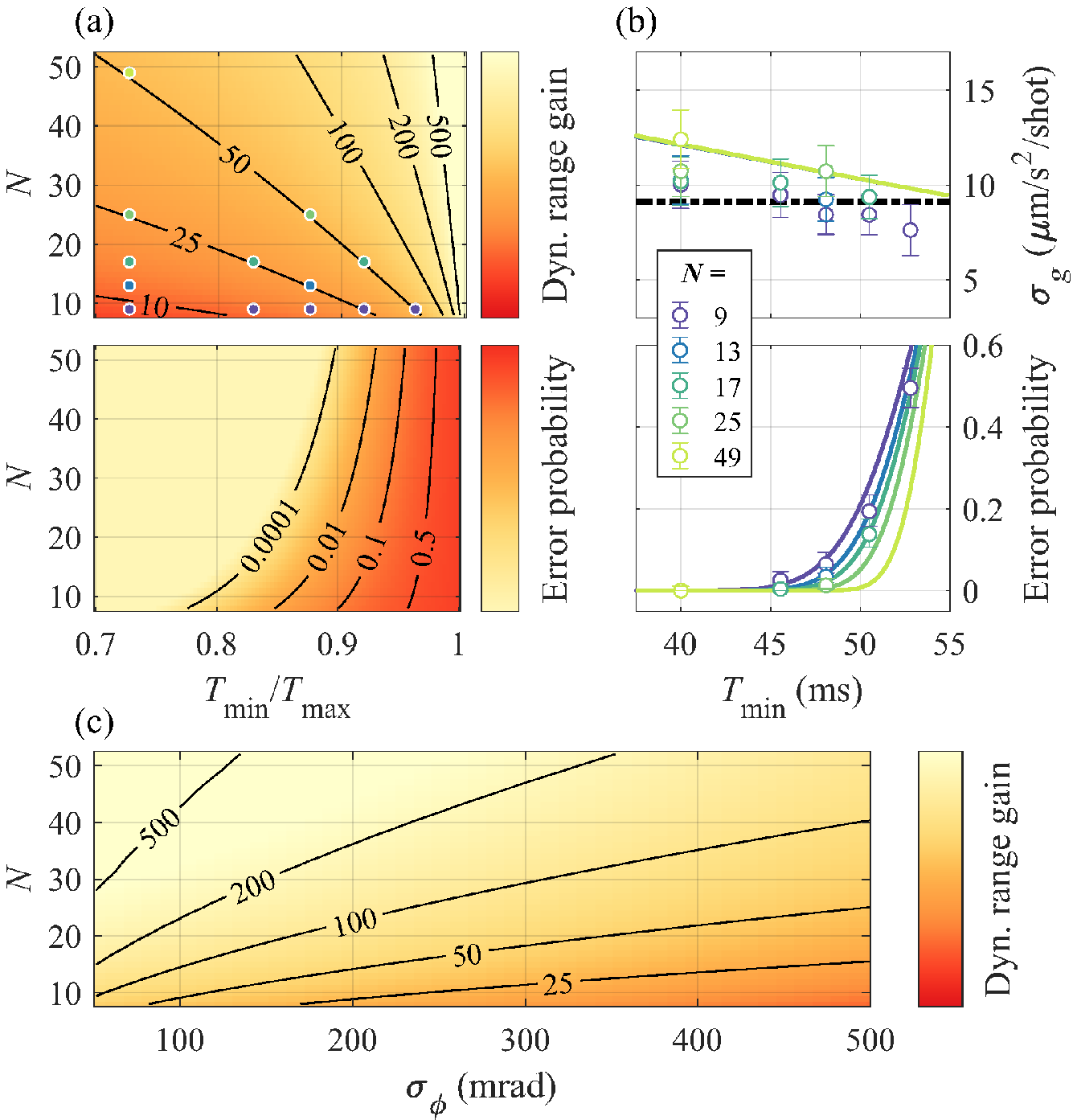}
\par\end{centering}
\centering{}\caption{(a) Projected performance of composite fringes compared to standard
measurement at fixed $T$. Top: Dynamic range enhancement {[}Eq.~\ref{eq:dyn-range-gain}{]}.
Bottom: Error probability {[}Eq.~\ref{eq:error-prob} in Appendix
A, evaluated with $\sigma_{\phi}=\unit[400]{mrad}${]}. (b) Experimental
results: Sensitivity \emph{per shot} (top) and error probability (bottom)
as a function of $T_{\min}$, with $T_{\max}=\unit[55]{ms}$, for
different $N$ as indicated in the legend. Measurements (circles)
are averaged over 230 composite fringes, error bars represent 68\%
confidence interval. The results are in excellent agreement with the
analytic predictions (solid lines). Dashed line represents a reference
measurement with fixed $T=\unit[55]{ms}$. Increase in dynamic range
at the measured settings is indicated by filled circles in (a). (c)
Projected gain in dynamic range as a function of $\sigma_{\phi}$
and $N$, for $10^{-2}$ error probability threshold.\label{fig:2-benchmark}}
\end{figure}

The potential gain in dynamic range is limited by $\sigma_{\phi}$.
Primarily, when the uncertainty in estimating the composite fringe
phase and frequency is comparable to the line separation in Fig.~\ref{fig:1-scheme}(d),
a \emph{line jump} may occur, resulting in a large error in estimating
$g$. The criterion for avoiding large errors is approximately (see
Appendix A)
\begin{equation}
\frac{\sqrt{N}}{\sigma_{\phi}}\frac{\left(T_{\textrm{max}}^{2}-T_{\textrm{min}}^{2}\right)}{T_{\textrm{min}}T_{\textrm{max}}}\gg1.\label{eq:error_condition}
\end{equation}
Notably, the potential gain in dynamic range increases if $\sigma_{\phi}$
decreases, as $T_{\textrm{min}}$ can approach $T_{\textrm{max}}$
without increasing the error probability. Conversely, the error probability
may be reduced by increasing $N$, which also serves to increase the
dynamic range, at the cost of temporal bandwidth. We note that in
the limit of $T_{\textrm{min}}=T_{\textrm{max}}$, Eq.~(\ref{eq:error_condition})
cannot be satisfied for finite $\sigma_{\phi}$, however such measurements
can still be analyzed as in traditional AI within the original dynamic
range $\Delta g_{\textrm{standard}}$, recovering the standard measurement
scheme at fixed $T$.

To understand the trade-offs in the composite fringe approach, we
present in Fig.~\ref{fig:2-benchmark}(a) the projected dynamic range
enhancement compared to fixed-$T$ measurements and the corresponding
error probability. The latter was evaluated for total phase uncertainty
$\sigma_{\phi}=\unit[400]{mrad}$, which was experimentally characterized
in our apparatus for $T\sim\unit[50]{ms}$ and is primarily due to
vibrations. Operating at $N\lesssim20$, corresponding to typical
AI temporal bandwidths, and requiring error probabilities below $10^{-2}$,
we can operate at $T_{\min}/T_{\max}\approx0.85$ and gain more than
an order of magnitude in dynamic range. For larger $N$, ratios $T_{\min}/T_{\max}$
of over $0.9$ become possible, enabling gains of over two orders
of magnitude.

To verify the model, we measured hundreds of composite fringes between
$T_{\min}=\unit[40]{ms}$ and $T_{\max}=\unit[55]{ms}$ over $N=49$
points at a constant chirp rate, and analyzed the results in subsets
of different $N$ and $T_{\min}$ values. Measurements at fixed $T=\unit[55]{ms}$
were also performed for reference. Figure~\ref{fig:2-benchmark}(b)
shows that the data are in excellent agreement with the analytical
model with no fit parameters.

\begin{figure*}[t]
\begin{centering}
\includegraphics[clip,width=2\columnwidth]{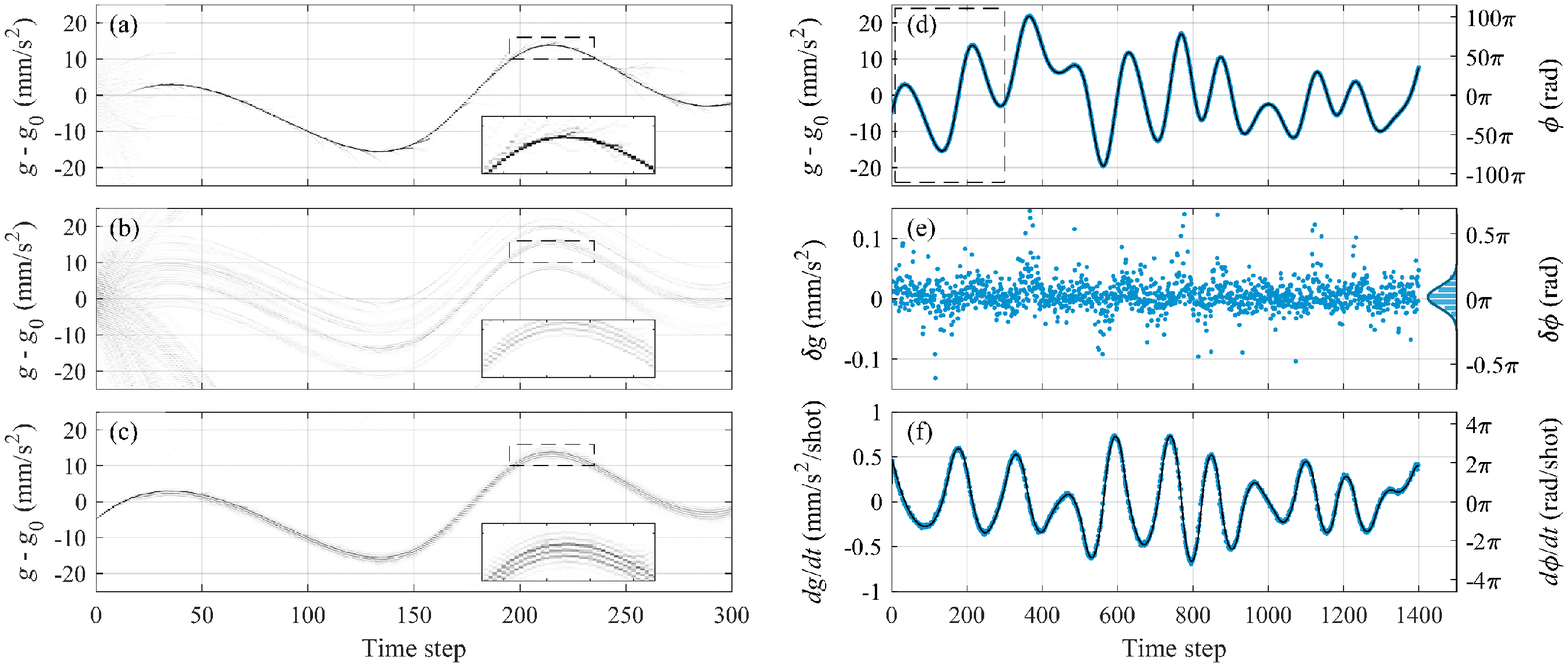}
\par\end{centering}
\centering{}\caption{Experimental results of composite fringe and particle filtering with
a dynamic signal. (a) Time-dependent, weighted particle distribution
for composite-fringe AI ($N=12,\thinspace T_{\min}=\unit[20]{ms},\thinspace T_{\max}=\unit[30]{ms}$,
corresponding to $\Delta g_{\textrm{comp}}=\unit[4.3]{mm/s^{2}}$).
Columns represent the weighted histogram of $g_{m,i}$ at each time
step. Particles are initialized at $g_{m,0}=g_{0}$, $\dot{g}_{m,0}=0$
with uncertainties $\sigma_{g,0}=\unit[5]{mm/s^{2}}$, $\sigma_{\dot{g},0}=\unit[1.5]{mm/s^{2}/\textrm{shot}}$.
(b) Particle filter analysis of standard fixed-$T$ measurements ($T=\unit[30]{ms}$),
same initial conditions as in (a). (c) Same as (b), with ideal initial
conditions and zero initial uncertainty. Insets depict the single
solution found by the filter using the composite-fringe approach versus
the ambiguity of standard fixed-$T$ measurements. (d) Estimated gravity
signal from our offline analysis protocol (blue; see Appendix B) compared
to the known input signal (black). Changes in $g$ correspond to $\unit[\pm100\pi]{rad}$
at $T_{\max}$, a hundred-fold improvement in dynamic range. Dashed
lines indicate the region shown in (a). (e) Estimated signal residuals,
with $\sigma_{g}=\unit[17]{\mu m/s^{2}}$ per shot, consistent with
the expected $\unit[15]{\mu m/s^{2}}$ per shot at these experimental
conditions. (f) $\dot{g}$ estimated by the filter (blue) compared
to the input signal (black). Shot-to-shot changes of up to $\unit[\pm3\pi]{rad/shot}$
at $T_{\max}$ are evident.\label{fig:3-particle-illustration}}
\end{figure*}

Figure~\ref{fig:2-benchmark}(c) summarizes the dynamic range enhancement
for different values of phase uncertainty, at a fixed error probability
threshold of $10^{-2}$, a value where large errors may be removed
with outlier detection with little sacrifice in sensitivity or bandwidth.
A large, realistic parameter space exists where the dynamic range
increases by more than $\times100$ at high temporal bandwidth.

\section{Time-varying signals}

The analysis thus far assumes a static or slowly varying signal with
respect to the measurement time of an $N$-point composite fringe,
as in stationary gravity measurements. We now turn our focus to dynamic
scenarios, such as mobile gravity surveys or inertial measurements
on navigating platforms, where the measured signal may change by more
than $\pm\pi$ from shot to shot and result in phase ambiguities.
To address this challenge and track a dynamic signal, we combine the
composite-fringe approach with an estimation protocol employing particle-filter
methodology.

Particle filtering is a Bayesian estimation protocol based on a sequential
Monte-Carlo method \citep{DelMoran1996,van2001unscented}, where a
large set of weighted particles is used to estimate the posterior
distribution of unknown state-variables based on inaccurate observations
or measurements. This approach is especially suited for problems with
multimodal likelihood functions, such as ambiguous phase measurements,
where the resulting (posterior) state distribution is very different
from Gaussian. Each time step of the filter consists of two actions:
First, particles propagate in state space according to an underlying
system model, forming a prediction of the new state. Second, a new
measurement is performed and particles are weighted according to their
likelihood given the current observed measurement.

In our particle filter implementation (see Appendix B), the state
variables are chosen as the instantaneous gravity value and its time
derivative. Each particle $m$ represents an hypothesis for these
variables at the $i$-th time step, defined as $g_{m,i}$ and $\dot{g}_{m,i}$.
The filter is initialized with equally-weighted, randomly drawn particles
characterized by initial uncertainties $\sigma_{g,0}$ and $\sigma_{\dot{g},0}$.
Prior to each measurement, particles are propagated according to $g_{m,i}=g_{m,i-1}+\dot{g}_{m,i-1}dt$
and $\dot{g}_{m,i}=\dot{g}_{m,i-1}+q_{m,i}$, where $q_{m,i}\sim\mathcal{N}\left(0,\sigma_{\dot{g}}^{2}\right)$
is a random process noise which allows the tracked signal to deviate
from purely linear behavior. Following each measurement, we update
the weight of each particle based on the likelihood that the measured
signal, \emph{i.e.}, $S_{i}=A-\left(C/2\right)\cos\phi_{i}$, is consistent
with the particle's hypothesized value $g_{m,i}$, assuming additive
white Gaussian detection noise. An estimate for $g_{i},\dot{g}_{i}$
is then derived from the weighted distribution of all particles.

We experimentally demonstrate the application of the particle filter
protocol to atom interferometry, both for standard fixed-$T$ measurements
and for the composite-fringe approach. By varying the Raman-lasers
chirp rate $\alpha$ between subsequent shots, we simulate an arbitrary
gravity signal with high frequencies and large amplitude, which may
arise, for example, in mobile gravimetry. Using the composite-fringe
approach, $T$ is varied over $N=12$ values between $T_{\min}=\unit[20]{ms}$
and $T_{\max}=\unit[30]{ms}$, for which $\sigma_{\phi}=\unit[140]{mrad}$
in our apparatus. To enhance the information gained at every measurement,
$T$ is sampled as $T_{1},T_{N/2+1},T_{2},T_{N/2+2},\ldots$ rather
than sequentially. As presented in Fig.~\ref{fig:3-particle-illustration}(a),
following an initial uncertainty period, the particles quickly converge
on the correct solution. New possible solutions occasionally emerge
but are quickly dismissed by the filter due to incompatibility with
incoming observations.

For comparison, Fig.~\ref{fig:3-particle-illustration}(b) presents
the same time-varying signal measured with fixed $T$. Here, many
trajectories remain likely as the filter is unable to converge on
the correct solution due to phase ambiguity. Finally, Fig.~\ref{fig:3-particle-illustration}(c)
shows the filter results with fixed $T$ when the initial conditions
of $g$ and $\dot{g}$ are assumed to be precisely known. Even in
this ideal and impractical scenario, fixed-$T$ measurements result
in ambiguities emerging over time and the exact solution for $g$
cannot be determined.

Figure~\ref{fig:3-particle-illustration}(d-f) presents complete
analysis of the measured data, using forward- and backward-propagating
particles to avoid edge effects (see Appendix B). The filter tracks
with high fidelity and high temporal resolution the varying signal,
which spans a dynamic range 100-times larger than a standard fringe
at $T_{\max}$ and which includes changes of more than $\unit[\pm3\pi]{rad/shot}$.
Additional improvement can be achieved by real-time estimation and
prediction of $g$, which would also allow tuning the interferometer
to mid-fringe at every measurement, as well as by incorporating quadrature
phase measurement \citep{Bonnin2018,Yankelev2019}.

\section{Conclusion}

In conclusion, we perform composite-fringe atom interferometry, employing
measurement sets with variable interrogation times. This approach
provides orders-of-magnitude gain in dynamic range by overcoming the
traditional trade-off between phase sensitivity and non-ambiguity
in interferometric sensors. An analytical model for the sensitivity
and error probability has been developed and compared to the experimental
study with excellent agreement.

Composite fringes are expected to benefit a range of applications
of atom interferometry. When measuring static or slowly varying signals,
as in gravity surveys where each survey point is measured for a short
period of time, traditional atom interferometry rely on spending a
substantial part of the sampling time measuring at low sensitivity
due to the large initial uncertainty of the measured signal. In contrast,
composite fringes allow for continuous high-sensitivity operation,
reducing the time necessary for reaching the target resolution in
each survey sample.

For dynamic scenarios, we have demonstrated the integration of a particle-filter
estimator for tracking signals which are impossible to measure with
traditional interferometry schemes without compromising sensitivity.
For applications such as gravity surveys on a continuously moving
platform, \emph{e.g.}, truck, ship or airplane, as well as for onboard
navigation applications, this approach will allow tracking signals
with larger variation than traditional interferometers, thereby lowering
the requirements from vibration isolation techniques or from post-processing
correlation with auxiliary sensors.

In many field applications, the need to measure inertial signals with
large dynamic range may be the limiting factor as it pushes the atom
interferometer to operate with short interrogation times and thus
at limited sensitivity. In these cases, the composite-fringe approach
significantly increases the possible sensitivity while maintaining
the necessary dynamic range.

Finally, while the discussion and experimental demonstration focused
on atomic gravimeters, this approach can be applied to other cold
atom sensors, including gravity gradiometers and gyroscopes, as well
as other interferometry-based quantum sensors such as magnetometers.

\section*{Acknowledgments}

We thank Ran Fischer, Roy Shaham, and Tal David for valuable discussions.
This work was supported by the Pazy Foundation and the Israel Science
Foundation.

\section*{Appendix A: Analytical model of composite fringe atom interferometry}

\renewcommand{\theequation}{A\arabic{equation}}
\setcounter{equation}{0}In this Appendix, we derive analytic expressions for the sensitivity
and error probability of atomic interferometry with composite fringes.
We begin with a composite fringe consisting of $N$ variable-$T$
measurements, 
\begin{equation}
S_{n}=A-\frac{C}{2}\cos\left[n\omega\left(g\right)+\phi\left(g\right)+\delta\phi_{n}\right]+\delta S_{n},\label{eq:varT_Sn_fringe-1}
\end{equation}
with $n=0,1,\ldots N-1$. The contrast $C$ and offset $A$ are assumed
to be known, and 
\begin{align}
\phi\left(g\right) & =\left(k_{\textrm{eff}}g-\alpha\right)T_{\textrm{min}}^{2},\label{eq:varphi}\\
\omega\left(g\right) & =\left(k_{\textrm{eff}}g-\alpha\right)\frac{T_{\textrm{max}}^{2}-T_{\textrm{min}}^{2}}{N-1},\label{eq:varomega}
\end{align}
are the unknown effective phase and frequency, as given in Eqs.~(\ref{eq:phi_g})-(\ref{eq:w_g})
of the main text. The fringe function explicitly includes the random
variables $\delta S_{n}$ and $\delta\phi_{n}$, representing realizations
of detection noise and phase noise, respectively, with variances $\sigma_{\textrm{det}}^{2}$
and $\sigma_{\textrm{phase}}^{2}$. The most dominant contribution
is usually phase noise due to mechanical vibrations of the mirror
retroreflecting the Raman beams, whose position sets the frame of
reference in which the atomic motion is measured. For small phase
noise, it is convenient to approximate the two noise terms by a single
effective detection noise $\delta\bar{S}_{n}$,

\begin{equation}
S_{n}\approx A-\frac{C}{2}\cos\left[n\omega\left(g\right)+\phi\left(g\right)\right]+\delta\bar{S}_{n},\label{eq:varT_Snstar}
\end{equation}
whose variance is given by
\begin{equation}
\bar{\sigma}^{2}=\left(\frac{1}{\sigma_{\textrm{det}}^{2}}+\frac{8}{C^{2}\sigma_{\textrm{phase}}^{2}}\right)^{-1}.
\end{equation}
The total phase uncertainty per shot, as defined in the main text,
is given by $\sigma_{\phi}=2\sqrt{2}\bar{\sigma}/C$. In the limit
of pure phase noise we have, as expected, $\sigma_{\phi}=\sigma_{\textrm{phase}}$.\\

\paragraph{(1) Sensitivity\protect \\
}

We examine the estimator $\boldsymbol{\theta}=\left[\phi,\omega\right]^{T}$
of the unknown parameters $\left[\phi\left(g\right),\omega\left(g\right)\right]^{T}$,
and we are interested in the uncertainty of $\boldsymbol{\theta}$,
as given by its covariance $\textrm{cov\ensuremath{\left(\boldsymbol{\theta}\right)}}$.
To this end, we adopt the framework of maximum likelihood estimation.
Given the set of measurements $S_{n}$, the likelihood function of
$\boldsymbol{\theta}$ is given by
\begin{align}
p\left(S_{n};\boldsymbol{\theta}\right) & =\frac{1}{\left(2\pi\bar{\sigma}^{2}\right)^{N/2}}\label{eq:pST}\\
 & \,\,\,\,\,\,\,\,\exp\left[-\frac{1}{2\bar{\sigma}^{2}}\sum_{n=0}^{N-1}\left(S_{n}-S\left(n,\boldsymbol{\theta}\right)\right)^{2}\right],\nonumber 
\end{align}
with $S\left(n,\boldsymbol{\theta}\right)=A-\left(C/2\right)\cos\left(n\omega+\phi\right).$
The Fisher information matrix is 
\begin{equation}
\left[I\left(\boldsymbol{\theta}\right)\right]_{i,j}=-E\left[\frac{\partial^{2}\ln p\left(S_{n};\boldsymbol{\theta}\right)}{\partial\theta_{i}\partial\theta_{j}}\right],\label{eq:EIP}
\end{equation}
where $E\left[\ldots\right]$ denotes expectation value. Substituting
Eq.~(\ref{eq:pST}) into Eq.~(\ref{eq:EIP}), we find
\begin{equation}
I\left(\boldsymbol{\theta}\right)=\left(\frac{C}{2\bar{\sigma}}\right)^{2}\frac{N\left(N-1\right)}{12}\left(\begin{array}{cc}
\frac{6}{N-1} & 3\\
3 & 2N-1
\end{array}\right),
\end{equation}
where we assumed large $N$ and that $\omega$ is not near $0$ or
$\pi$ (the latter can be relaxed by deliberately varying $\phi_{L}$
when measuring a composite fringe). For an unbiased estimator of $\boldsymbol{\theta}$,
the Cramér-Rao lower bound (CRLB) of the covariance matrix is given
by $\textrm{cov\ensuremath{\left(\boldsymbol{\theta}\right)}}\ge I\left(\boldsymbol{\theta}\right)^{-1}$,
so

\begin{equation}
\textrm{cov\ensuremath{\left(\boldsymbol{\theta}\right)}}\ge\left(\frac{\bar{\sigma}}{C}\right)^{2}\frac{16}{N\left(N+1\right)}\left(\begin{array}{cc}
2\left(N-1\right) & -3\\
-3 & \frac{6}{N-1}
\end{array}\right).\label{eq:cov}
\end{equation}

Given the best estimated phase and frequency $\tilde{\boldsymbol{\theta}}$,
we now search for an estimator $\tilde{g}=\tilde{g}(\tilde{\boldsymbol{\theta}})$
for the actual acceleration $g$. The optimal estimator is the value
of $\tilde{g}$ that minimizes the product $[\boldsymbol{\theta}\left(\tilde{g}\right)-\tilde{\boldsymbol{\theta}}]^{T}I\left(\boldsymbol{\theta}\right)[\boldsymbol{\theta}\left(\tilde{g}\right)-\tilde{\boldsymbol{\theta}}]$,
where $\boldsymbol{\theta}(\tilde{g})=\left[\phi\left(\tilde{g}\right),\omega\left(\tilde{g}\right)\right]^{T}$,
with $\phi\left(\tilde{g}\right)$ and $\omega\left(\tilde{g}\right)$
given in Eqs.~(\ref{eq:varphi}-\ref{eq:varomega}). Performing the
minimization for different values of $\tilde{\boldsymbol{\theta}}$
yields the function $\tilde{g}=\tilde{g}(\tilde{\boldsymbol{\theta}})$.
With this function in hand, the uncertainty in estimating $g$ is
found by the transformation
\begin{equation}
\sigma_{g}^{2}\ge\left[\frac{\partial\tilde{g}(\tilde{\boldsymbol{\theta}})}{\partial\tilde{\boldsymbol{\theta}}}\right]^{T}\textrm{cov\ensuremath{\left(\boldsymbol{\theta}\right)}}\left[\frac{\partial\tilde{g}(\tilde{\boldsymbol{\theta}})}{\partial\tilde{\boldsymbol{\theta}}}\right].\label{eq:cov_transformation}
\end{equation}
Using Eq.~(\ref{eq:cov}), we find 
\begin{align}
\sigma_{g} & \ge\frac{\sqrt{8}\bar{\sigma}}{Ck_{\mathrm{eff}}T_{\max}T_{\min}}\label{eq:error-prob}\\
 & \cdot\left[1+\frac{1}{6}\frac{2N-1}{N-1}\left(\frac{T_{\min}}{T_{\max}}\frac{T_{\max}^{2}-T_{\min}^{2}}{T_{\min}^{2}}\right)^{2}\right]^{-1/2},\nonumber 
\end{align}
which, to leading order in $\left(T_{\max}^{2}-T_{\min}^{2}\right)/T_{\min}^{2}$
and in terms of the total phase uncertainty $\sigma_{\phi}=2\sqrt{2}\bar{\sigma}/C$,
simplifies to

\begin{equation}
\sigma_{g}\gtrsim\frac{\sigma_{\phi}}{k_{\mathrm{eff}}T_{\max}T_{\min}},
\end{equation}
as given in the main text.\\

\paragraph{(2) Large estimation error\protect \\
}

We now turn to calculate the probability of a large estimation error,
which results from a \emph{jump} between the lines in the phase map
{[}Fig.~\ref{fig:1-scheme}(d){]}. First, we derive the uncertainty
of the estimated phase and frequency along an axis perpendicular to
these lines. The coordinate along this axis is defined as $p=-\phi\sin\alpha+\omega\cos\alpha$,
with $\tan\alpha=\left(T_{\textrm{max}}^{2}-T_{\textrm{min}}^{2}\right)/\left[T_{\textrm{min}}^{2}\left(N-1\right)\right]$.
The uncertainty in $p$ is then given by a transformation similar
to that in Eq.~(\ref{eq:cov_transformation}) and found to be

\begin{align}
\sigma_{p} & =4\frac{C}{\bar{\sigma}}\\
 & \cdot\sqrt{\frac{6\left(N-1\right)T_{\textrm{min}}^{2}T_{\textrm{max}}^{2}+\left(2N-1\right)\left(T_{\textrm{max}}^{2}-T_{\textrm{min}}^{2}\right)^{2}}{N\left(N+1\right)[\left(N-1\right)^{2}T_{\textrm{min}}^{4}+\left(T_{\textrm{max}}^{2}-T_{\textrm{min}}^{2}\right)^{2}]}}\nonumber 
\end{align}
or, to leading order in $\left(T_{\max}^{2}-T_{\min}^{2}\right)/T_{\min}^{2}$,

\begin{equation}
\sigma_{p}\approx\frac{C}{\bar{\sigma}}\frac{T_{\max}}{T_{\min}}\sqrt{\frac{96}{N\left(N^{2}-1\right)}}.
\end{equation}
On the other hand, the distance between the lines is given by
\begin{equation}
\Delta p=\frac{2\pi\left(T_{\textrm{max}}^{2}-T_{\textrm{min}}^{2}\right)}{\sqrt{\left(N-1\right)^{2}T_{\textrm{min}}^{4}+\left(T_{\textrm{max}}^{2}-T_{\textrm{min}}^{2}\right)^{2}}}.
\end{equation}
From this we find the ratio, again to leading order in $\left(T_{\max}^{2}-T_{\min}^{2}\right)/T_{\min}^{2}$
and in terms of the total phase uncertainty $\sigma_{\phi}$,
\begin{equation}
\frac{\Delta p}{2\sigma_{p}}=\frac{\pi}{\sqrt{12}}\frac{1}{\sigma_{\phi}}\frac{\left(T_{\textrm{max}}^{2}-T_{\textrm{min}}^{2}\right)}{T_{\textrm{min}}T_{\textrm{max}}}\sqrt{N\frac{N+1}{N-1}}.
\end{equation}
The probability of a large estimation error in $g$ is then
\begin{equation}
\epsilon=2\left[1-\Phi\left(\frac{\Delta p}{2\sigma_{p}}\right)\right],
\end{equation}
where $\Phi\left(x\right)$ is the normal cumulative distribution
function.

\section*{Appendix B: Particle-filter implementation}

\renewcommand{\theequation}{B\arabic{equation}}
\setcounter{equation}{0}In our implementation of the particle filter, we use the state variables
$g$ and $\dot{g}$ to describe the dynamic system. The state of the
$m$th particle at the $i$th time step is thus defined as 
\begin{equation}
\boldsymbol{x}_{m,i}=\left[\begin{array}{c}
g_{m,i}\\
\dot{g}_{m,i}
\end{array}\right].
\end{equation}
The propagation model is given by $\boldsymbol{x}_{m,i+1}=\mathbf{F}\cdot\boldsymbol{x}_{m,i}+\boldsymbol{w}_{m,i}$,
where the state propagation matrix $\mathbf{F}$ is
\begin{equation}
\mathbf{F}=\left[\begin{array}{cc}
1 & dt\\
0 & 1
\end{array}\right],
\end{equation}
and $\boldsymbol{w}_{m,i}$ is a random process noise, distributed
normally with zero mean and with a covariance given by
\begin{equation}
\mathbf{Q}=dt^{2}\left[\begin{array}{cc}
0 & 0\\
0 & \sigma_{\dot{g}}^{2}
\end{array}\right],
\end{equation}
where $dt$ is the time increment between two measurements.

The input to the filter is the interferometer signal $S_{i}$, measured
at each time step with a different interrogation time $T_{i}$. The
filter also receives as input the interferometer fringe parameters
$A_{i}$, $C_{i}$, which are found separately by collecting all measurements
of the same interrogation time $T_{i}$ and fitting their distribution.
At each time step and for each particle, the residual is calculated
as
\begin{equation}
r_{m,i}=S_{i}-\left[A_{i}-\frac{C_{i}}{2}\cos\left(k_{\textrm{eff}}g_{m,i}T_{i}^{2}\right)\right],
\end{equation}
from which the likelihood $p\left(\boldsymbol{x}_{m,i}|S_{i}\right)$
is determined based on the measurement noise model,
\begin{equation}
p\left(\boldsymbol{x}_{m,i}|S_{i}\right)=\frac{1}{\sqrt{2\pi\bar{\sigma}^{2}}}\exp\left[-\frac{r_{m,i}^{2}}{2\bar{\sigma}^{2}}\right].\label{eq:p_likelihood}
\end{equation}
Each particle is weighted according to its likelihood, and all particles
are finally resampled at every time step with systematic resampling.
Eq.~(\ref{eq:p_likelihood}) may be generalized for non-stationary
detection noise, by including time-step dependent $\bar{\sigma}_{i}$,
as well as for noise with non-Gaussian distribution by changing the
functional form itself.

We used 5000 particles in the examples presented in the main text.
Traditionally, at the end of each time step, the state variables are
estimated as a weighted mean of all particles. We achieve more stable
results by running the filter both forward and backward in time, calculating
the time-dependent histogram of particles from both directions together,
and running a ridge-detection algorithm (MATLAB \texttt{tfridge} function)
on the combined histogram to find a continuous estimation of $\tilde{g}_{i}$.
This analysis is less sensitive to temporary branching of the particles
distribution.

A distribution of residuals of the measurement data can be calculated
with respect to the estimated measurements, \emph{i.e.},
\begin{equation}
\tilde{r}_{i}=S_{i}-\left[A_{i}-\frac{C_{i}}{2}\cos\left(k_{\textrm{eff}}\tilde{g}_{i}T_{i}^{2}\right)\right].
\end{equation}
The parameter $\sigma_{\dot{g}}$ is determined by minimizing the
variance of the $\tilde{r}_{i}$.

\bibliography{Variable_T_and_Particle_Filter}

\end{document}